\documentclass[pra,aps,twocolumn,showpacs,superscriptaddress,amsmath,amssymb]{revtex4-1}
\usepackage{color}

\usepackage{amsmath,graphicx}

\begin{document}
\def\tr{\rm{Tr}}
\def\la{{\langle}}
\def\ra{{\rangle}}
\def\a{{\alpha}}
\def\p{\tilde{p}}
\def\x{\tilde{x}}
\def\w{\tilde{W}}
\def\t{\tilde{t}}
\def\k{\tilde{k}}
\def\s{\tilde{\sigma}}
\def\l{\hat{\Lambda}}
\def\a{\hat{A}}
\def\n{\hat{n}}

\title{Atomic Fock states by gradual trap reduction:  from sudden to adiabatic limits}
%
%
\author {D. Sokolovski}
\affiliation{Departmento de Qu\'imica-F\'isica, Universidad del Pa\' is Vasco, UPV/EHU, Leioa, Spain}
\affiliation{IKERBASQUE, Basque Foundation for Science, E-48011 Bilbao, Spain}
\affiliation{School of Maths and Physics, Queen's University of Belfast, Belfast, BT7 1NN, UK}
\author{M. Pons}
\affiliation{Departmento de F\'isica Aplicada I, EUITMOP, Universidad del Pa\' is Vasco, UPV/EHU, Barakaldo, Spain}
\author{A. del Campo}
\affiliation{Institut f{\"u}r Theoretische Physik, Albert-Einstein Allee 11, Universit{\"a}t Ulm, D-89069 Ulm, Germany}
\author{J. G. Muga}
\affiliation{Departmento de Qu\'imica-F\'isica, Universidad del Pa\' is Vasco, UPV/EHU, Leioa, Spain}

\date{\today}
\begin{abstract}
We investigate the possibility to form high fidelity atomic Fock states by gradual reduction of a quasi one dimensional trap containing spin polarized fermions or strongly interacting bosons in the Tonk-Girardeau regime.
Making the trap shallower and simultaneously squeezing it can lead to the preparation of an ideal atomic Fock state  as one approaches either the sudden or the adiabatic limits. Nonetheless, the fidelity of the resulting state is shown to exhibit a non-monotonic behaviour with the time scale in which the trapping potential is changed.

\end{abstract}

%
%
\pacs{PACS numbers: 37.10.Gh, 03.75.Kk, 05.30.Jp}
\maketitle
\section{Introduction }
Preparation of atomic states containing exactly a fixed number $M$ of atoms (Fock states) is of importance for a wide range of applications, from studying ultracold chemistry and few-body physics to precision measurements and quantum information processing. The aim of such a preparation is to create a quantum state with the mean number of atoms $\la n\ra$ equal to the chosen $M$ and its mean variance as small as possible.
Different proposal based on a time-dependent modulation of the confining potential have been put forward both for optical traps \cite{DRN07,MUG,PONS,WRN09,RWZN09} and optical lattices \cite{Qdistill, NP10}.

Recent experiments have shown the possibility of achieving atom-number sub-Poissonian statistics
in a quantum degenerate gas with repulsive interactions \cite{CSMHPR05}.
The key idea is that the mean number of trapped atoms can be controlled by adiabatically reducing the depth of the trap
while expelling the excess of atoms. The precision of the control improves by maximising the energy splitting between
states with different number of particles which ultimately allows one to discriminate such states by modulating the depth of the trapping potential.
This technique is referred as to atom culling \cite{DRN07}. For ultracold gases confined in tight-waveguides,
it works optimally in the strongly interacting regime for Bosonic samples \cite{DRN07,MUG,PONS,WRN09},
i. e., in the Tonks-Girardeau gas, where the repulsive interactions lead to
an effective Pauli exclusion principle \cite{Girardeau60}.
Alternatively, Fock state preparation is optimized as well with a spin-polarized non-interacting Fermi gas \cite{PONS,RWZN09}.
 Bearing in mind that both systems (which are dual under Bose-Fermi mapping \cite{Girardeau60}) share all local correlation functions,
 and in particular the atom-number distribution \cite{Moll}, we shall address the polarized fermions for brevity in the following.
Nonetheless, all results in this paper apply to both systems.

Non-interacting spinless fermions when placed in a trap occupy the lowest one-particle levels in the ground state.
Changing the shape of the potential trap aims to expel redundant atoms into continuum levels leaving the bound states of the final trap
filled to its maximum capacity, $\la \n\ra = M$.
This condition together with the Pauli exclusion principle ensures high fidelity of the preparation:
since no more than $M$ atoms can be distributed over $M$ levels of the final well,
the variance $\sigma^2=\la \n^2\ra -\la \n\ra^2$ must vanish, and precisely $M$ atoms will be
found in each individual case.
A slow change of the trapping potential would guarantee, by virtue of the adiabatic theorem,
full occupation of the states in the final well. The adiabaticity may, however, require times
 large compared to the time scales typically involved in ultracold atom experiments and one might wish for
 a faster way to achieve full final occupation.
One of the counter intuitive results obtained in Refs.  \cite{MUG, PONS} is that an infinitely fast
``sudden'' change provides an alternative to the  adiabatic route which can lead to the preparation of ideal Fock states, when
making the well shallower (weakening of the trap) is accompanied by also making it narrower (squeezing).
Further, at zero temperature, the best results are obtained  \cite{MUG} in case the projector on the subspace
of all filled initial states $\l_0$ contains the projector $\l$
on the subspace spanned by all bound states of the final well ($\l\subseteq\l_0 $),
i.e., when  on that subspace $\l_0$ can be approximated by unity.
In addition, non-zero temperature effects can be overcome by starting with a larger initial sample \cite{PONS}.
\newline
However, the sudden limit may not be accessible in a given experimental set up,
and the main purpose of the present paper is to explore the behaviour of the trapped atoms
when a change of the potential is neither sudden, nor adiabatically slow.
In particular, we will show that the efficiency of the Fock state preparation exhibits
a non-monotonic behaviour as a function of the duration of the quench of the trapping potential.
We will also address other questions such as the dependence of final occupation
on the depths
of the initial and the final wells, and the escape time required for the expelled atoms to leave the well region.
The rest of the paper is organised as follows. In Sect. II we will study the dependence of
the final occupation on the time in which the trapping potential changes its shape.
In Sect. III we briefly discuss the behaviour of the variance when the final well is filled close to its
maximal capacity. In Sections IV we review two different mechanisms responsible for the formation of nearly ideal Fock states in the adiabatic and sudden limits.
In Sections V we discuss the escape of expelled particles from the
final well region,
and Section VI contains our conclusions.

\section{Atom counting statistics following a change of the trapping potential}
Consider  a number of non-interacting spinless fermions (fermionised bosons) initially trapped in a one-dimesional  rectangular
square well of depth $V_i$ and width $L_i$. The potential $V(x,t)$ undergoes, over a  time $T$, a transformation such that its final shape is also a rectangular well, but with new
 depth and width, $V_f$ and $L_f$, respectively. Although other types of evolution are possible, we will
 consider linear change of both the depth and width of the trap,
\begin{eqnarray}\label{a1}
V(t)=V_i+(V_f-V_i)t/T, \\
\nonumber
L(t)=L_i+(L_f-L_i)t/T,
\end{eqnarray}
where the ramping time $T$ determines whether evolution of the potential is rapid or slow.
Our aim is to control the mean number $\la n(T) \ra$ of the fermions trapped in the final trap, while minimizing the variance
 \begin{eqnarray}\label{a2}
\sigma^2_N(T) \equiv \la n^2(T)\ra-\la n(T)\ra^2,
\end{eqnarray}
for the case of zero temperature.
As already mentioned in the introduction, non-zero temperature effects  turn out to be non-critical and can be conveniently overcome by starting with a larger initial sample as described in \cite{PONS}.
We will further assume that the initial and final traps support $K$ and $M \le K$ bound states, respectively, and that there are $N\le K$ fermions occupying the first $N$ levels of the initial well.
A well known technique \cite{Lev,Klich,Moll} allows to express
$\la n(T) \ra$ and $\sigma^2_N(T)$ for a system of non-interacting  fermions in terms of the solutions of the corresponding one-particle Schr\"odinger equation.
Let us denote by $\phi^i(x)$ ($\phi^f(x)$) the states of the initial (final) trap.
Bound (scattering) states will be labeled by a Latin (Greek) letter as a subindex.
In the dimensionless variables $x/L_i$ and $t/t_0$, where
\begin{eqnarray}\label{a3}
t_0\equiv \mu L_i^2 / \hbar ,
\end{eqnarray}
and $\mu$ is the atomic mass, the time-dependent single-particle eigenstates obey
\begin{eqnarray}\label{a4}
i\partial_t \phi_n^i =-\frac{1}{2}\partial_x^2 \phi_n^i +W(x,t)\phi_n^i,\quad n=1,\dots,N,
\end{eqnarray}
with $W(x,t)=V(x,t)t_0/\hbar$.
For a trap size of $80$ $\mu$m  the time $t_0$ for Rb and Cs atoms
takes values of $8.8$ s and $13.4$ s, respectively.
Following \cite{MUG,PONS,Lev,Klich} we obtain
\begin{eqnarray}\label{1}
\la n(T)\ra =\sum_{j=1}^M \la \phi_j^f|\l_T|\phi_j^f\ra
\end{eqnarray}
and
\begin{eqnarray}\label{2}
\sigma^2_N(T) = \la n(T)\ra
-\sum_{j=1}^M\la \phi_j^f| \l_T \l \l_T)|\phi_j^f\ra.
\end{eqnarray}
%
Here, $\l$ is the projector on the subspace spanned by the one particle bound states
$|\phi^f_j\ra$, $j=1,2,\dots,M$ of the final well,
\begin{eqnarray}\label{4}
\l = \sum_{j=1}^M |\phi_j^f\ra \la \phi_j^f|.
\end{eqnarray}
Similarly, $\l_T$ is the projector onto the subspace spanned by the orthogonal states
obtained by the time evolution of the one particle bound states $|\phi^i_n\ra$, $n=1,2,\dots,N$, in the initial well,
\begin{eqnarray}\label{3}
\l_T = \sum_{n=1}^N |\phi^T_n\ra \la \phi^T_n|, \quad |\phi^T_n\ra  \equiv \hat{U}(T)  |\phi^i_n\ra,
\end{eqnarray}
with $\hat{U}(T)$ denoting the evolution operator corresponding to Eq. (\ref{a4}).
Note that Eqs. (\ref{1}) and (\ref{2}) are generalisations of Eqs. (9) and (10) obtained in Ref. \cite{MUG}
for the sudden limit, with the initial bound states replaced by time evolved states (\ref{3}).
We further notice that knowledge of time evolved states (\ref{3}) allows one to
obtain the full atom-number distribution $p(n)$ from the characteristic function $F(\theta)=\tr[\hat{\rho}e^{i\theta\hat{\Lambda}\hat{n}\hat{\Lambda}}]$,
as a Fourier transform, $p(n)=\frac{1}{2\pi}\int_{-\pi}^{\pi}e^{-in\theta}F(\theta)d\theta$, with $n=1,\dots,M$ \cite{Klich,Moll,Lev}.
Using the projector for the bound subspace in the final configuration, $F(\theta)={\rm det}{\bf A}$ with ${\bf A}=[1+(e^{i\theta}-1)\hat{\Lambda}\hat{\Lambda}_T]$.
In the basis of single-particle eigenstates $|\phi_m^f\ra$, the elements of the matrix ${\bf A}$ read $A_{nm}=\delta_{nm}+[\exp(i\theta)-1]\la\phi_n^f|\hat{\Lambda}_T|\phi_m^f\ra$.

With the problem reduced to numerical evaluation of the corresponding one particle states,
we employ the Crank-Nicolson method to solve Eq. (\ref{a4}) with the initial conditions
$\phi_n^i(x,t=0)$
$n=1,2...N$ and zero boundary conditions at the edges of the numerical box,
$x=\pm L_b$.
To avoid reflections from the boundaries we introduce an absorbing potential
proposed by Manolopoulos \cite{Mano,MPNE04}  (see appendix A for details).
Following Refs. \cite{MUG,PONS}, we will
refer as ``weakening'' to the case where the trap is made shallower while keeping its width constant,
$V_i>V_f$, $L_i=L_f$ , while reducing the trap's width will be called ``squeezing''.

From Eqs. (\ref{1}) and  (\ref{2}) it is easy to deduce \cite{MUG, PONS}
that an ideal Fock state with $\la n(T)\ra=M$ and $\sigma^2_N(T)=0$ would be prepared provided that $\l\subseteq\l_T$,
this is, the space spanned by the state before the end of the quench should enclose that spanned by the Fock state to be prepared \cite{proof}.
In the following we shall describe different physical implementations to fulfill this requirement.
\section{Two ways to arrive at a good Fock state}
%
%
%
%
%
\begin{figure}[h]
\includegraphics[width=8cm, angle=0]{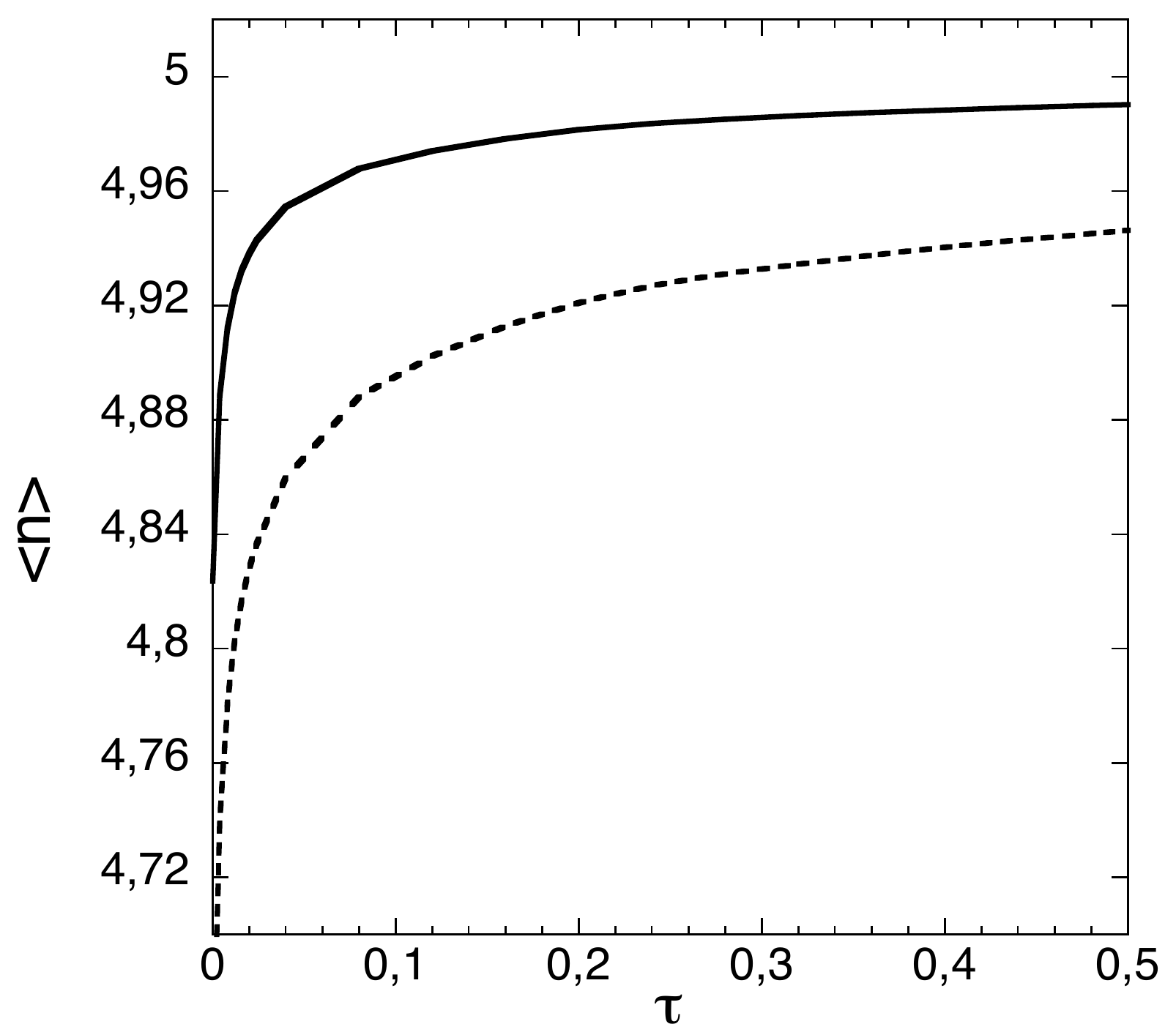}
\caption{Dependence of the final occupation on the ramping time (pure weakening):
$\la n(T)\ra$ (solid) and $\la n(T)\ra -\sigma_N(T)$ (dashed) vs $\tau=T/t_0$.
The potentials are $W_i=(19.5)^2\pi^2/2$, $W_f=(4.95)^2\pi^2/2$, corresponding to
maximum capacities $K=20$ and $M=5$, respectively, and $N=K=20$, it is, all initial levels occupied.\\
}
\label{FIG1}
\end{figure}
In the {\it  adiabatic limit}, where the change of the potential shape is slow, we can expect the first
$M$ bound states of the initial well to follow into the $M$ bounds states of the final well
\begin{eqnarray}\label{b1a}
|\phi^T_n\ra= \exp(i\Phi_n) |\phi_n^f\ra, \quad n=1,2,\dots,M,
\end{eqnarray}
where $\Phi_n$ is a real phase.
Thus, on the subspace of final bound states we have
\begin{eqnarray}\label{b2}
\l_T =   \sum_{n=1}^M |\phi^T_n\ra\la \phi^T_n| =  \sum_{n=1}^M |\phi^f_n\ra\la \phi^f_n| .
\end{eqnarray}

Figure \ref{FIG1} shows the final occupation $\la n(T)\ra$ and its variance as functions
of the ramping time for the case of pure weakening of the trap.
As expected, the final occupation increases and the fidelity improves as one approaches the adiabatic limit, where the variance of the resulting state vanishes.
This result (a consequence of the adiabatic theorem) does not depend on whether weakening, squeezing or a combination of both
techniques is applied.

It is worth noting that,
to our knowledge, the question of exactly how slowly the potential must evolve to ensure adiabaticity in the case the
$M$-th level in the final well lies close to the edge of continuum
remains open and requires further investigation. As mentioned in the introduction,
the applicability of the adiabatic method may be limited by a finite lifetime of the trapped condensate.

In the {\it sudden limit} where the shape of the well changes almost instantaneously
we have $\hat{U}(T)\approx 1,$
$|\phi_{\nu}^T\ra= |\phi_{\nu}^i\ra$ and $|\phi_{n}^T\ra= |\phi_{n}^i\ra$, i. e. $\l_T=\l_0$.
Let us recall that if $\l\subseteq\l_T$, i.e, the final well would contain a Fock state with precisely $M$ fermions.
Remarkably, the use of either either weakening and squeezing in the sudden limit lead to
quasi-Fock states of limited fidelity \cite{MUG}, while the combination of both techniques can lead to ideal Fock states \cite{MUG,PONS}.
Generally, sudden quenches of the trap might lead to undesirable excitations of the transverse modes breakdown the effective one-dimensional character of the system.
\section{Gradual modulation of the trap potential}
We have seen that both adiabatic and sudden changes of the trapping potential can in principle
lead to the preparation of ideal Fock states,
but as idealized limits might be of limited relevance to experimental implementation of atom culling techniques.  

\begin{figure}[h]
\includegraphics[width=8cm, angle=0]{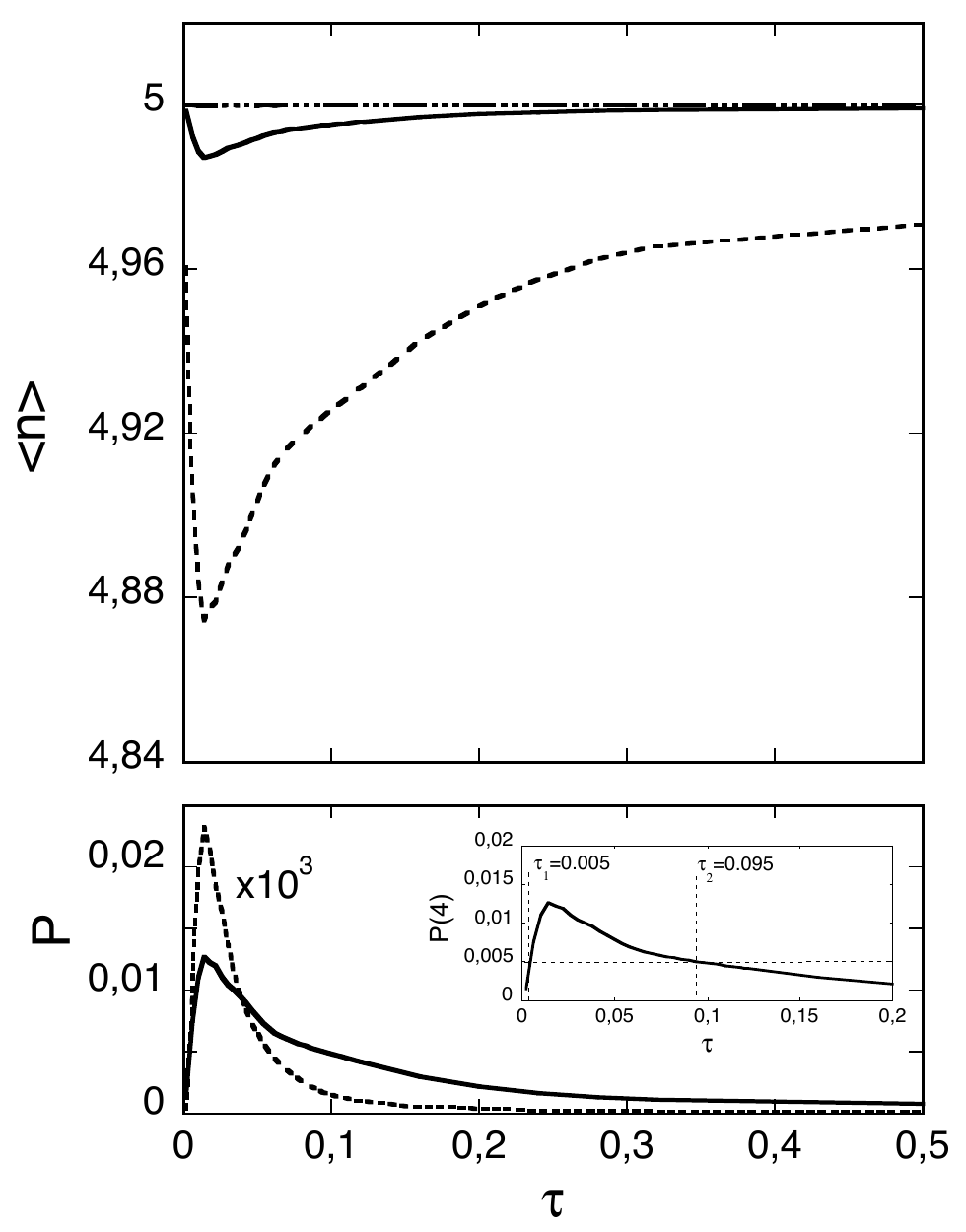}
\caption{a). Same as Fig. \ref{FIG1} but for weakening combined with squeezing, $L_f/L_i=0.6$.
Also shown by a dot-dashed line is the l.h.s. of Eq. (\ref{11}).\\
b) Probabilities $p(4)$ (solid)  and $p(3)\times 10^3$ (dashed) in Eqs. (\ref{11a}) vs $\tau=T/t_0$. The inset shows the probability $p(4)$ and the ramping time interval for which $p(4)>0.5\%$.}
\label{FIG2}
\end{figure}

Motivated by this observation we next look at the mean particle number and variance of a
state resulting from a combined process of squeezing and weakening of the trap in finite time, see
Fig. \ref{FIG2}.
Here,  $\la n(T)\ra$ stays close to the maximum capacity of the well for all ramping times, but exhibits  a dip at $T/t_0 \approx 0.03-0.1$.
Notice that the adiabatic preparation relies on the adiabatic following of the first $M$ states of the initial trap onto the $M$
states of the bound subspace of the final trap \cite{DRN07}. By contrast, the sudden preparation relies on the instantaneous resolution of the target state $|M\ra$
within the space spanned by the initial state \cite{MUG,PONS}.
The non-monotonic behaviour of $\la n(T)\ra$ reported in Fig. \ref{FIG2} results from a simultaneous failure of these two different mechanisms
that allow the perfect resolution of the desired Fock state in either of the $T\rightarrow 0,\infty$ limits.

We further note (see Fig. 2, dashed) that in the case of combined weakening and squeezing the
mean number of atoms and its variance (both dimensionless, of course) add almost exactly to the maximum
capacity of the final well, $M$, for all ramping times, and in the next section we discuss this approximate ``sum rule'' in some detail.
\section{Sub-Poissonian statistics at nearly full final occupation}
Let us start by introducing the deficiency operator
\begin{eqnarray}\label{7}
\hat{A}=\l-\l_T.
\end{eqnarray}


Inserting (\ref{7}) into (\ref{1}) and  (\ref{2}) yields
\begin{eqnarray}\label{8}
\la n(T)\ra =M-\sum_{j=1}^M \la \phi_j^f|\a|\phi_j^f\ra
\end{eqnarray}
and
\begin{eqnarray}\label{9}
\sigma^2_N(T) = \sum_{j=1}^M \la \phi_j^f|\a|\phi_j^f\ra -
 \sum_{j,k=1}^M|\la \phi_j^f|\hat{A}|\phi_k^f\ra|^2.
\end{eqnarray}
From Eqs. (\ref{8}) and (\ref{9})  follows  a ``sum rule''
\begin{eqnarray}\label{10}
\la n(T)\ra + \sigma^2_N(T) = M -  \sum_{j,k=1}^M
|\la \phi_j^f|\hat{A}|\phi_k^f\ra|^2.
\end{eqnarray}
Note that the last term is quadratic in the operator $\hat{A}=1_f-\l_T$ so that  if $\a$ is ``small'' \cite{FOOT}  the variance can be expressed in terms of $\la n(T)\ra$ through the
approximate relation
\begin{eqnarray} \label{11}
\la n(T)\ra + \sigma^2_N(T) \approx M,
\end{eqnarray}
as demonstrated in Fig. 2a.
This is an example of non-Poissonian behaviour resulting from
indistinguishability of the fermions involved. Indeed, for a Poissonian distribution one would
expect  $\sigma^2_N(T) = \la n(T)\ra$ whereas from Eq. (\ref{11}) we have
$\sigma^2_N(T) = M- \la n(T)\ra \ll \la n(T)\ra$.

Note also that in the case of almost full final occupation, knowing
$\la n(T)\ra$ and $\sigma^2_N(T)$ allows one to reconstruct the full counting statistics.
Thus, neglecting the probabilities to trap less than $M- 2$ fermions, $p(k)\approx 0$
for $k=0,1,\dots,M-3$, we can obtain $p(M-2)$, $p(M-1)$ and $p(M)$ from the normalisation and the  first two moments of the distribution $p$,
\begin{eqnarray} \label{11a}
\sum_{k=M-2}^M p(k) k^m= \la n ^m \ra, \quad m=0,1,2.
\end{eqnarray}
Results of such a reconstruction are shown in Fig. 2b.
For the set corresponding to Fig. \ref{FIG2}b the probability to trap three fermions, $p(3)$,
is negligibly small, and one would obtain $4$ fermions at most in $1.2\%$ of all cases for
$\tau=T/t_0\approx 0.01$. To reduce this fraction to $0.5\%$ one has an option to choose the ramping time
either less than $\tau_1 < 0.005$ or longer than $\tau_2> 0.095$ shown on the inset in Fig. \ref{FIG2}b.
%
%
%
%

%
%
%
\section{Escape of expelled particles from the final trap region}
\begin{figure}[h]
\includegraphics[width=7cm, angle=0]{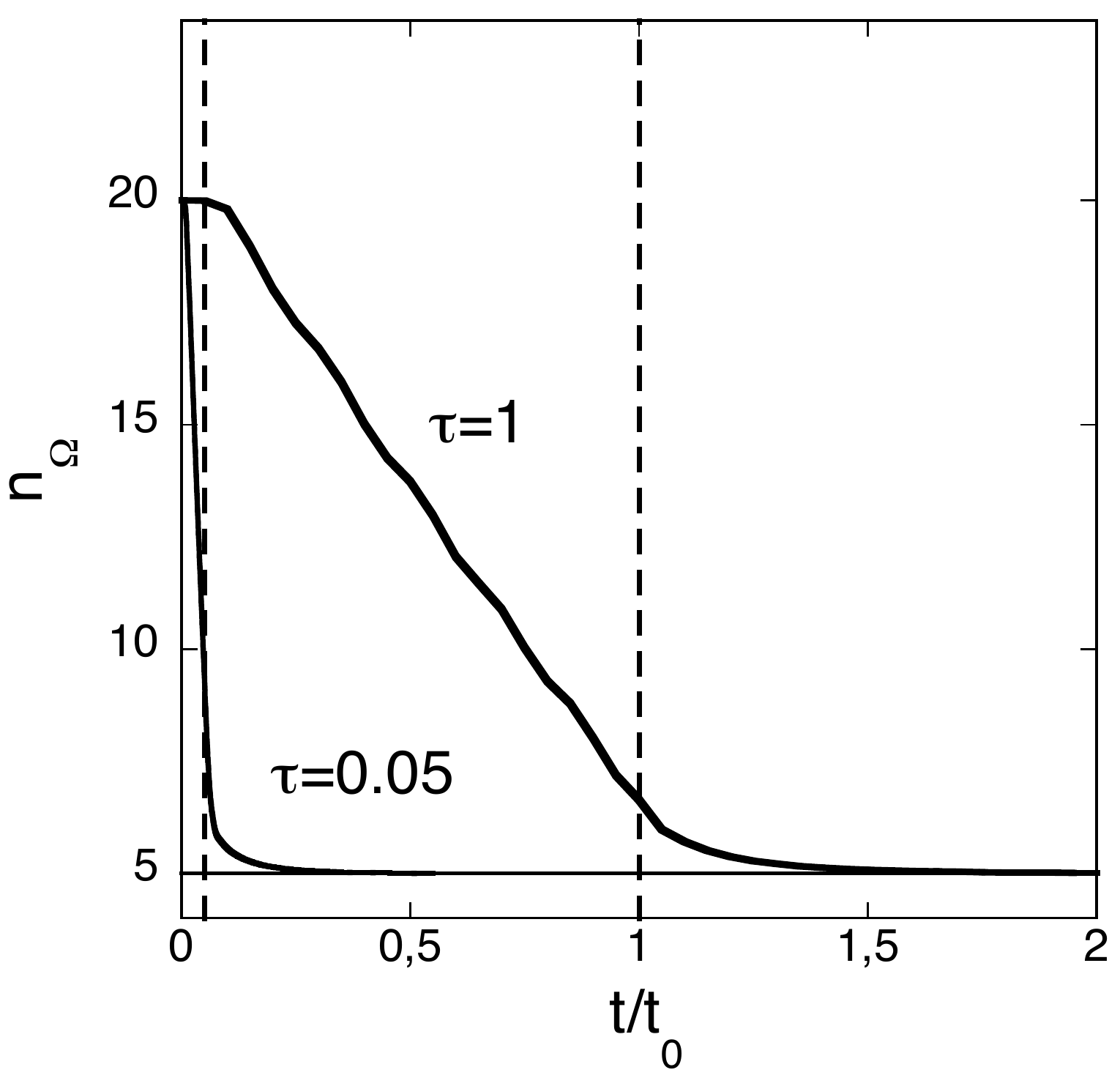}
\caption{Mean number of atoms in well region $(\Omega:$ $-0.75 L_f < x < 0.75 L_f)$ vs
$t/t_0$ for $\tau=T/t_0=0.05$ and $\tau=T/t_0=1$. $W_i=(19.5)^2\pi^2/2$, $N=K$
and weakening is combined with squeezing, $L_f/L_i =0.6$. Vertical dashed lines indicate
ramping times after which the potential assumes its final form.\\}
\label{FIG6}
\end{figure}

Although the final occupation of the bound states in the final well is determined
by the time $T$ after which the potential no longer changes, the obtained Fock state can only be used after the atoms expelled into the continuum have left the well area. To obtain an estimate  for
how fast this would happen we have chosen an interval $\Omega$: $-1.5 L_f \le x \le 1.5 L_f$, containing the final well region, and monitored the mean number of atoms in $\Omega$, $\la n_{\Omega}(t)\ra$, during the ramping, $t\le T$, as well as for $t>T$. Figure \ref{FIG6} shows the results for combined
weakening and squeezing, $W_f/W_i = 0.18$, $L_f/L_i = 0.6.$  for $T/t_0=0.05$ and
$T/t_0=1.0$, close to the sudden and the adiabatic limits, respectively.
With this choice of parameters, the initial well contains $20$ atoms and the final well's maximum
capacity is $5$.

In the nearly sudden limit, $T/t_0=0.05$, a considerable fraction of expelled atoms remain in the region $\Omega$ by the time the well achieves its final configuration. One has then to wait a duration of the order of $t_0$ for the mean number of atoms in $\Omega$ to settle to its final value
$\la n(T)\ra \approx M =5$. A more detailed analysis shows that it is
the $M+1$ (in this case, the sixth) state of the initial well which is delayed most in leaving the area.
We note also that this decay is not exponential and, therefore, cannot be attributed solely to trapping of an atom in one of the resonances supported between the edges of the final well.

In the nearly adiabatic case, $T/t_0=1$, initial one-particle levels are gradually pushed into the continuum, and most of the atoms have time to leave $\Omega$ while the potential shape is still changing.
Once the change has stopped one still has to wait approximately $t_0$ for the contribution from
the $M+1$-th state to clear the area. Although comparable in magnitude, the wait is somewhat longer than that in the nearly sudden limit, mostly because the expelled atoms receive more energy  and move faster if the change of the potential shape is sudden.
%
%
%
%
\section{Conclusions and discussion}
In summary, we have investigated formation of atomic Fock states by gradually changing the trapping potential and expelling the excess of atoms out of the trap.
For the case in which either the depth (weakening) or the width (squeezing) of the potential trap are reduced,
an increase of the ramping time leads to an improvement in the fidelity of the procedure until the adiabatic limit is reached.
This monotonic behaviour in a relevant variable is generic in phenomena with sudden to adiabatic crossovers. For example, in a Landau-Zener transition in which the crossing time is gradually increased, the final state goes from one level to the other depending on that time and the corresponding degree of adiabaticity of the passage. However, in our case,  
a combination of both procedures (squeezing and weakening), which produces
high fidelity Fock states in the sudden limit, remains superior to either of both operations for all ramping times. This opens a new route for feasible atomic Fock state creation by trap reduction, approximating the sudden limit in the laboratory. 
The behaviour along the crossover is non monotonic and for finite quenching times one observes a decrease in the mean number of trapped atoms
which is accompanied by a corresponding increase in its variance.
Therefore in this combined scheme we find the rather unusual property that 
both sudden and adiabatic limits lead to the same results, certainly through different mechanisms, with a non-montonic decrease of fidelity in between.      
\section{Acknowledgement}
We acknowledge support of University of Basque Country UPV-EHU
(Grant GIU07/40), Basque Government (IT-472-10), and  Ministry of
Science and Innovation of Spain (FIS2009-12773-C02-01). JGM and AdC acknowledge the hospitality of the Max Planck Institute for the Physics of Complex Systems at Dresden, Germany. 
%

\appendix

\section{Complex absorbing potential}
The absorbing potential employed in this paper has the form suggested in \cite{Mano,MPNE04}.
With the wavefunction required to vanish at the edges of the computational box, $x=\pm L_b$ denoting the size,
the absorbing potential is chosen to be zero for  $0 <x<L_b/2$. For
$L_b/2 \le x< L_b$  it is given by
\begin{eqnarray}\label{5}
V_{abs}(x)= -iD\bigg[Az-Bz^3+\frac{4}{(C-z)^2}-\frac{4}{(C+z)^2}\bigg],\nonumber\\
\end{eqnarray}
where $z=C(2x/L_b-1)$, $C=2.62206$, $A=(1-16/C^3)$, $B=(1-17/C^3)/C^2$, and
$D=C^2L_b^2/1.28$. Finally, $V_{abs}(x)=V_{abs}(-x)$.
The value $L_b/L_i=10$ was used and the Schr\"odinger equation was solved numerically
on a grid of $4\cdot 10^5$ points spanning the interval $[-L_b,L_b]$.

\section{Final occupation vs the depth of  the initial trap}

\begin{figure}[h]
\includegraphics[width=7cm, angle=0]{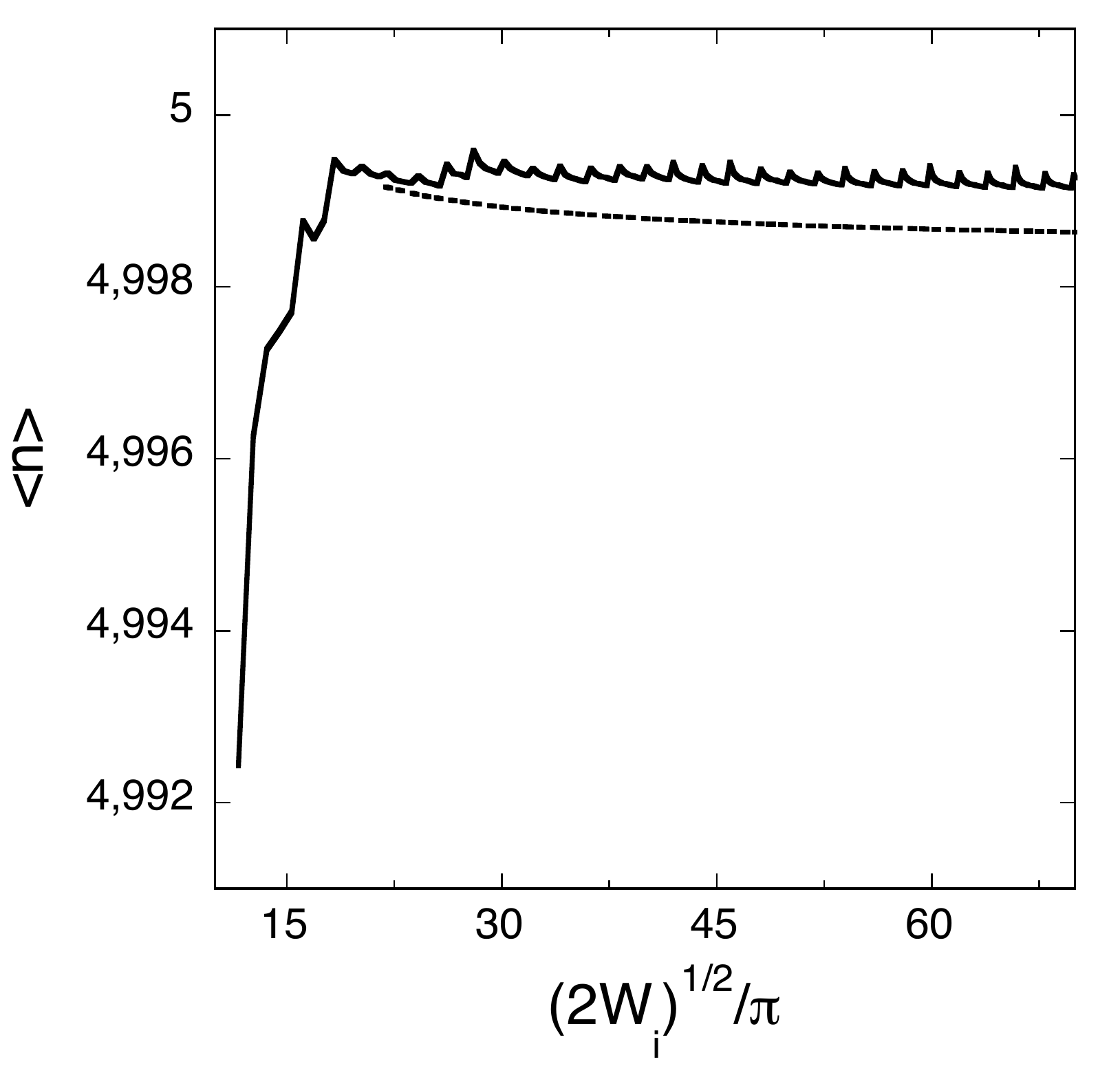}
\caption{Final occupation in the sudden limit, $\la n(T=0)\ra$, for the case of weakening with squeezing vs. the dimensionless initial well depth
for $N=K$ (solid) and $N=25$ (dashed). Other parameters are $W_f=(4.95)^2\pi^2/{2(L_f/L_i)^2}$ ($M=5$) and $L_f/L_i=0.6$. Note that for a deep well  $(2W)^{1/2}/\pi$ gives an estimate of the number of bound states,  $K \approx Int ((2W)^{1/2}/\pi)$.
}.
\label{FIG3}
\end{figure}

The discussion in Section IV suggests that, close to the sudden limit and for weakening and squeezing applied together, starting with a deeper well fully filled,
$N=K$, should improve fidelity. In addition, starting with a deeper initial well with a fixed number of level filled, $N$, would not be as beneficial, since the frequency of oscillations of the
first unfilled state, $\la x|\phi_{N+1}^i\ra$,  is almost independent of $V_i$.
 This is illustrated in Fig. \ref{FIG3} where the solid line shows the final occupation
of the well with the maximum capacity $M=5$ with all the levels of the initial well filled, $N=K$.
The graph show peaks which correlate with initial depths at which a new level appears in the initial well. In the same figure, the dashed line shows the dependence of $\la n(T)\ra$ on $V_i$ when only first $25$ levels of the initial well are occupied. In this case, $\la n(T)\ra$  is no longer sensitive to the appearance of new bound states, but yields a slightly lower fidelity of the preparation.

\begin{figure}[h]
\includegraphics[width=7cm, angle=0]{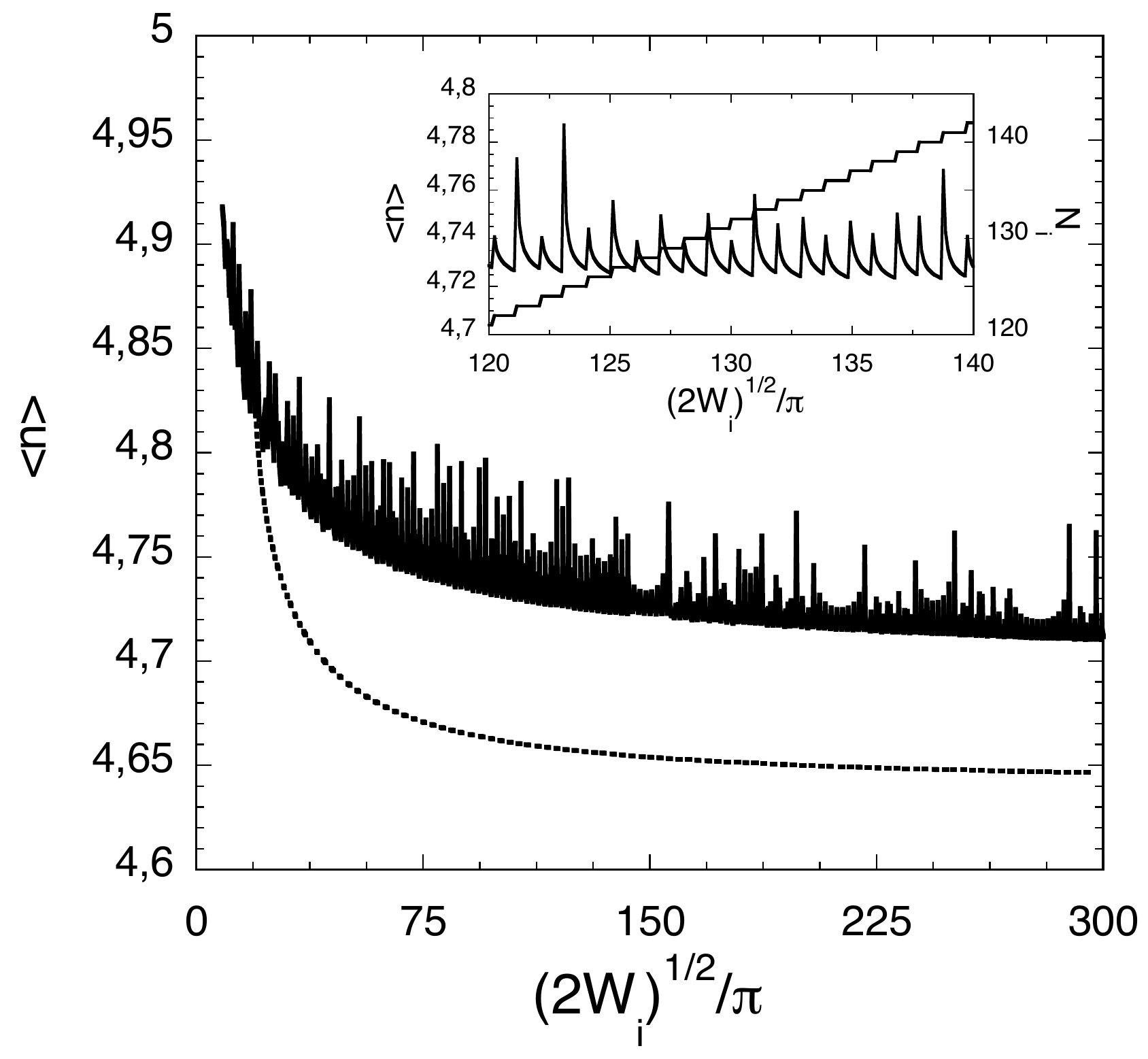}
\caption{Same as Fig. \ref{FIG3} but for the case of pure weakening,  $L_i=L_f$
and for $N=K$ (solid) and $N=25$ (dashed). The insert shows the correlation between the structure in the $N=K$ curve
and the number of the bound states supported by the initial well.
}.
\label{FIG4}
\end{figure}

The case of pure weakening shown in Fig. \ref{FIG4}  is more complex.
 There increasing the depth of a fully filled well leads to a decrease in
 $\la n(T)\ra$ (solid), which is also highly sensitive to the increase in the number of initial bound states
 (see inset in Fig. \ref{FIG4}). As in Fig. \ref{FIG3}, partial filling of the initial well removes the structure in the dependence of  $\la n(T)\ra$ on $V_i$ but leads to lower final occupation in the deep well limit.


\begin{thebibliography}{999}
\bibitem{proof} Whenever $\hat{\Lambda}\subseteq\hat{\Lambda}_T$,
$A_{nm}=\exp(i\theta)\delta_{nm}$, $F(\theta)=\exp(i\theta M)$ and the
the desired Fock space is prepared: $p(n)=\delta_{n,M}$.
\bibitem{FOOT}
More precisely, we need both the sum, $I_1$,  in Eq. (\ref{8}) and the double sum, $I_2$, in Eq. (\ref{9}) to be
small compared to unity. Denoting  $A_{max}$ the largest of $|\la \phi_j^f|\hat{A}|\phi_k^f\ra|$, with
$j,k=1,2,\dots,M$ we find $I_1\le MA_{max}$ and $I_2\le M^2A_{max}^2$.
Thus, for the contribution of the operator $\a$ to be negligible, $MA_{max} \ll 1$.
\end{thebibliography}
\end{document}